\documentstyle[epsfig]{mn}

\newif\ifAMStwofonts

\ifoldfss
  \ifCUPmtlplainloaded \else
    \NewTextAlphabet{textbfit} {cmbxti10} {}
    \NewTextAlphabet{textbfss} {cmssbx10} {}
    \NewMathAlphabet{mathbfit} {cmbxti10} {} 
    \NewMathAlphabet{mathbfss} {cmssbx10} {} 
  \fi
  \ifAMStwofonts
    \ifCUPmtlplainloaded \else
      \NewSymbolFont{upmath} {eurm10}
      \NewSymbolFont{AMSa} {msam10}
      \NewMathSymbol{\upi}     {0}{upmath}{19}
      \NewMathSymbol{\umu}     {0}{upmath}{16}
      \NewMathSymbol{\upartial}{0}{upmath}{40}
      \NewMathSymbol{\leqslant}{3}{AMSa}{36}
      \NewMathSymbol{\geqslant}{3}{AMSa}{3E}

       \let\le=\leqslant
       \let\ge=\geqslant
    \fi
  \fi
\fi 

\ifnfssone
  \newmathalphabet{\mathit}
  \addtoversion{normal}{\mathit}{cmr}{m}{it}
  \addtoversion{bold}{\mathit}{cmr}{bx}{it}
  \newmathalphabet{\mathbfit} 
  \addtoversion{normal}{\mathbfit}{cmr}{bx}{it}
  \addtoversion{bold}{\mathbfit}{cmr}{bx}{it}
  \newmathalphabet{\mathbfss} 
  \addtoversion{normal}{\mathbfss}{cmss}{bx}{n}
  \addtoversion{bold}{\mathbfss}{cmss}{bx}{n}
  \ifAMStwofonts
    \ifCUPmtlplainloaded \else
      \UseAMStwoboldmath
      \makeatletter
      \new@mathgroup\upmath@group
      \define@mathgroup\mv@normal\upmath@group{eur}{m}{n}
      \define@mathgroup\mv@bold\upmath@group{eur}{b}{n}
      \edef\UPM{\hexnumber\upmath@group}
      \new@mathgroup\amsa@group
      \define@mathgroup\mv@normal\amsa@group{msa}{m}{n}
      \define@mathgroup\mv@bold\amsa@group{msa}{m}{n}
      \edef\AMSa{\hexnumber\amsa@group}
      \makeatother
      \mathchardef\upi="0\UPM19
      \mathchardef\umu="0\UPM16
      \mathchardef\upartial="0\UPM40
      \mathchardef\leqslant="3\AMSa36
      \mathchardef\geqslant="3\AMSa3E

       \let\le=\leqslant
       \let\ge=\geqslant
    \fi
  \fi
\fi 

\ifnfsstwo
  \DeclareMathAlphabet{\mathbfit}{OT1}{cmr}{bx}{it}
  \SetMathAlphabet\mathbfit{bold}{OT1}{cmr}{bx}{it}
  \DeclareMathAlphabet{\mathbfss}{OT1}{cmss}{bx}{n}
  \SetMathAlphabet\mathbfss{bold}{OT1}{cmss}{bx}{n}
  \ifAMStwofonts
    \ifCUPmtlplainloaded \else
      \DeclareSymbolFont{UPM}{U}{eur}{m}{n}
      \SetSymbolFont{UPM}{bold}{U}{eur}{b}{n}
      \DeclareSymbolFont{AMSa}{U}{msa}{m}{n}
      \DeclareMathSymbol{\upi}{0}{UPM}{"19}
      \DeclareMathSymbol{\umu}{0}{UPM}{"16}
      \DeclareMathSymbol{\upartial}{0}{UPM}{"40}
      \DeclareMathSymbol{\leqslant}{3}{AMSa}{"36}
      \DeclareMathSymbol{\geqslant}{3}{AMSa}{"3E}

       \let\le=\leqslant
       \let\ge=\geqslant
    \fi
  \fi
\fi 

\ifCUPmtlplainloaded \else
  \ifAMStwofonts \else 
    \def\upi{\pi}
    \def\umu{\mu}
    \def\upartial{\partial}
  \fi
\fi

\title{Accretion onto black holes formed by direct collapse}

\author[J.L. Johnson et al.]
{Jarrett L. Johnson$^1$\thanks{E-mail: jjohnson@mpe.mpg.de}, Sadegh Khochfar$^1$, Thomas H. Greif$^2$, Fabrice Durier$^1$ \\
$^1$Max-Planck-Institut f{\"u}r extraterrestrische Physik, Giessenbachstra\ss{}e, 85748 Garching, Germany \\
$^2$Max-Planck-Institut f{\"u}r Astrophysik, Karl-Schwarzschild-Stra\ss{}e 1, 85740, Garching, Germany}

\pubyear{2010}

\begin{document}
\maketitle
\topmargin-1cm

\begin{abstract}
One possible scenario for the formation of massive black holes (BHs) in the early Universe is 
from the direct collapse of primordial gas in atomic-cooling dark matter haloes in which 
the gas is unable to cool efficiently via molecular transitions.  We study the formation of 
such BHs, as well as the accretion of gas onto these objects and the high energy radiation 
emitted in the accretion process, by carrying out cosmological radiation hydrodynamics 
simulations.  In the absence of radiative feedback, we find an upper limit to the accretion rate onto the central object which forms from the 
initial collapse of hot ($\la$ 10$^4$ K) gas of the order of 0.1 
M$_{\odot}$ yr$^{-1}$.  This is high enough for the formation of a supermassive star, the immediate precursor of a BH, with a mass 
of the order of 10$^5$ M$_{\odot}$.  Assuming that a fraction of this mass goes into a BH, we track the subsequent accretion of gas onto the BH 
self-consistently with the high energy radiation emitted from the accretion disk.  Using a 
ray-tracing algorithm to follow the propagation of ionizing radiation, we  
model in detail the growth and evolution of the H~{\sc ii} and He~{\sc iii} regions which 
form around the accreting BH.  We find that BHs with masses of the order of 10$^4$ M$_{\odot}$ 
initially accrete at close to the Eddington limit, but that the accretion rate drops 
to $\sim$ 10$^{-5}$ M$_{\odot}$ yr$^{-1}$ (of order 1 percent of the Eddington limit) 
after $\sim$ 10$^6$ yr, due to the expansion of the gas near the 
BH in response to strong photoheating and radiation pressure.  One distinctinve signature of the accretion of gas onto 
BHs formed by direct collapse, as opposed to massive Pop III star formation, is an extremely high 
ratio of the luminosity emitted in He~{\sc ii} $\lambda$1640 to that emitted in H$\alpha$ (or Ly$\alpha$), i.e. $L_{\rm 1640}$ / 
$L_{\rm H\alpha}$ $\ge$ 2; this nebular emission could be detected by future facilities, such as the {\it James Webb Space Telescope}.  
Finally, we briefly discuss implications for the coevolution of BHs and their host galaxies.     
\end{abstract}


\section{Introduction}
How did the seeds for the first supermassive black holes (BHs) form?  Observations of high redshift quasars have shown that BHs can grow rapidly in the early Universe, becoming as 
massive as $\ga$ 10$^9$ M$_{\odot}$ within the first billion years of cosmic time (e.g. Willott et al. 2003; Fan et al. 2004; 2006; see also the recent reviews Haiman 2009; Volonteri 2010).
However, it is still an open question how such BHs can grow to be so massive in such a short time.  Given the strong connection between galaxies and the BHs that
grow in the centers of their host dark matter (DM) haloes (e.g. Magorrian et al. 1998; Richstone et al. 1998; Silk \& Rees 1998; Ferrarese \& Merritt 2000), 
the growth of the first massive black holes must be intimately related to the formation of the first galaxies (e.g. Bromm et al. 2009). 
While the galaxies hosting the first massive black holes at redshifts $z$ $\ga$ 10 have not yet been probed by observations, the radiation generated by massive black holes 
accreting in the early Universe will be the target of upcoming telescopes operating at various wavelengths, 
such as the {\it Energetic X-ray Imaging Survey Telesope}\footnote{http://exist.gsfc.nasa.gov/} (e.g. Grindlay 2005; Grindlay et al. 2010), 
the {\it James Webb Space Telescope}\footnote{http://www.jwst.nasa.gov/} (JWST; e.g. Gardner et al. 2006), 
and the {\it Atacama Large Millimeter Array}\footnote{http://www.eso.org/sci/facilities/alma/} (see e.g. Schleicher et al. 2010a).
Therefore, in addition to the question of how the first black holes acquire their mass, it is
important and timely to also consider the observational signatures of accretion onto the first massive BHs and whether they will be observable with future facilities.

A growing body of theoretical work has focused on the question of how supermassive BH seeds are formed and grow in the early Universe, 
either from the collapse of Population (Pop) III stars (e.g. Haiman \& Loeb 2001; Yoo \& Miralda-Escud{\' e} 2004; Volonteri \& Rees 2005; Johnson \& Bromm 2007; Alvarez et al. 2009; 
Tanaka \& Haiman 2009; see also Devecchi \& Volonteri 2009) or from the direct collapse of primordial gas in atomic-cooling haloes (e.g. Bromm \& Loeb 2003; Begelman 2006; 
Spaans \& Silk 2006; Lodato \& Natarajan 2006; Regan \& Haehnelt 2009a,b; Shang et al. 2010; Schleicher et al. 2010b). 
In the former scenario, a BH forms with an initial mass of the order of 100 M$_{\odot}$, or roughly the mass of its Pop III progenitor; in the latter case, the initial BH mass is likely
$\ga$ 10$^4$ M$_{\odot}$, thereby providing a substantial jump start towards its growth to $\sim$ 10$^9$ M$_{\odot}$ by $z$ $\sim$ 6.

The accretion rate and associated radiative feedback from the growth of BHs in the early Univsere, in particular, has been studied by numerous authors 
by carrying out cosmological simulations (see e.g. Li et al. 2007; Pelupessy et al. 2007; Di Matteo et al. 2008; Sijacki et al. 2009; Booth \& Schaye 
2009; Levine et al. 2010; see also Sutter \& Ricker 2010).  Such simulations have commonly used simple prescriptions to estimate the accretion rate onto BHs and to gauge the 
effect of the radiation emitted during the accretion process (e.g. by injecting thermal energy without a radiative transfer calculation).  
Recently, however, complementary theoretical investigations have treated the radiative feedback from BH accretion on BH growth in great detail, 
but have used idealized, non-cosmological initial conditions (e.g. Milosavljevi{\'c} 2009a,b; Park \& Ricotti 2010). 

In a study complementary to previous cosmological simulations concerning the accretion onto BHs formed from Pop III stars, 
which has been shown to be ineffecient due to radiative feedback both from the preceding Pop III star (e.g. Yoshida 2006; Johnson \& Bromm 2007; but see Whalen et al. 2008) 
and from the accretion of gas onto the BH itself (Alvarez et al. 2009), here we consider the accretion of gas onto BHs formed by direct collapse 
in an atomic-cooling cosmological halo with a virial temperature of $\sim$ 10$^4$ K at $z$ $\sim$ 15, in which H$_2$ is strongly 
suppressed by a uniform background photodissociating radiation field. 
 We employ a ray-tracing scheme to track the propagation of H~{\sc i}- and He~{\sc ii}-ionizing photons, to account for the 
photodissociation of H$_2$ molecules, and to include the effects of radiation pressure on the primordial gas in the vicinity of the BH.  
Our simulations self-consistently couple the accretion rate of primordial gas onto the BH to the emission of high energy 
radiation from the accretion disk, the spectrum of which is calculated as a function of the BH mass and accretion rate at every timestep.  
Our results allow us to accurately calculate the rate of accretion of gas onto the BH and, importantly, to determine 
the observable signatures of the radiation emitted in the accretion process as it is reprocessed into nebular emission which escapes the host halo.   
 
In the next Section, we describe the setup of our cosmological simulation and the evolution of the primordial gas as it collapses into an atomic-cooling DM halo.  
In Section 3, we discuss the accretion of gas onto the central collapsed object that forms at the center of such a halo, prior to its collapse to form a BH. 
Our calculation of the rate of accretion of gas onto such a newly formed BH and our treatment of the radiation emitted in the accretion process is described in Section 4. We present
our results on the accretion rate and associated radiative feedback in Section 5, we consider the observational signatures of such BH accretion in the earliest galaxies in Section 6, and
we explore the implications of our results for the coevolution of BHs and their host galaxies in Section 7.  
Finally, in Section 8 we conclude with a brief summary and discussion.

\begin{figure}
\vspace{2pt}
\epsfig{file=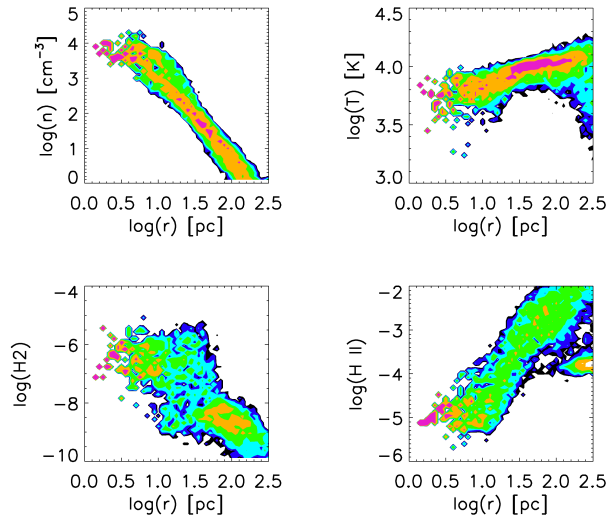,width=8.4cm}
\caption{The properties of the primordial gas exposed to a molecule-dissociating background radation field with $J_{\rm LW}$ = 10$^3$, at $z$ $\sim$ 15, as a function of distance from the most dense SPH particle, at the location of which a black hole is expected to form by direct collapse.  Clockwise, from top-left: the number density of hydrogen nuclei, the gas temperature, the H~{\sc ii} fraction, and the H$_2$ fraction.  Contours denote the distribution of the gas at a fixed radius $r$, the mass fraction doubling across contour lines.  These properties are in rough agreement with those that previous authors have found in similar simulations (see Wise et al. 2008; Regan \& Haehnelt 2009b; Shang et al. 2010).  Note that due to the generally low H$_2$ fraction, the gas temperature remains of the order of a few times 10$^3$ K even at high densities in the core of the halo.  
}
\end{figure}

\section{Collapse of primordial gas in an atomic-cooling halo}
The formation of a BH by direct collapse is predicted to occur when a sufficiently large (e.g. $\ga$ 10$^5$ M$_{\odot}$) mass of primordial gas is unable to cool 
much below $\sim$ 10$^4$ K via molecular transitions and accumulates at the center of a DM halo with a comparable virial temperature 
(e.g. Bromm \& Loeb 2003; Dijkstra et al. 2008; Shang et al. 2010).  Our simulation setup is chosen to realize these conditions.
As is common in this scenario for direct BH formation, we invoke a strong Lyman-Werner (LW) background radiation field in order to suppress 
cooling of the primordial gas by H$_2$; however, we note that, in principle, other heating sources could also act to keep the gas sufficiently hot 
(e.g. Sethi et al. 2010; see also Schleicher et al. 2010c).

\subsection{Simulation Setup}
We carry out our three-dimensional numerical simulations with the parallel version of GADGET 
(version 1), which includes a tree (hierarchical) gravity solver combined with the smoothed particle hydrodynamics 
(SPH) method for tracking the evolution of gas (Springel et al. 2001; Springel \& Hernquist 2002).  Along with 
H$_2$, H$_2^{+}$, H, H$^-$, H$^+$, e$^-$, He, He$^{+}$, and He$^{++}$, we have included the five deuterium species 
D, D$^+$, D$^-$, HD and HD$^+$, using the same chemical network as in Johnson \& Bromm (2006, 2007).

\begin{figure}
\vspace{2pt}
\epsfig{file=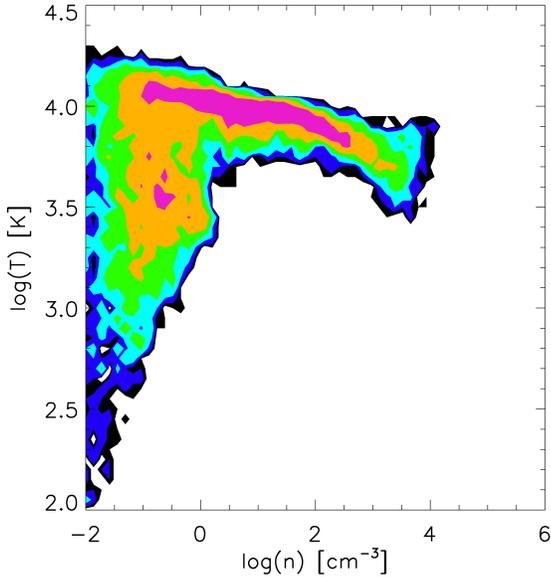,width=8.4cm}
\caption{The temperature of the gas within 1 kpc of the center of the most massive halo in our simulation volume, as a function of the hydrogen number density, just before a sink particle representing the central massive collapsed object is put in place at $z$ $\sim$ 15.  Here contours denote the relative mass fraction of all of the gas shown (not of the gas at fixed radius, as in Fig. 1), this fraction doubling across contour lines.  The virial temperature of the 3.5 $\times$ 10$^7$ M$_{\odot}$ host halo is 1.8 $\times$ 10$^4$ K.  As the contours show at $n$ $\sim$ 10 cm$^{-3}$, most of the gas collapsing to higher density in the halo is shock heated 
near the virial radius ($r_{\rm vir}$ $\sim$ 600 pc) to $\sim$ 10$^4$ K, thereafter cooling by atomic transitions at higher densities.
}
\end{figure}

\begin{figure}
\vspace{2pt}
\epsfig{file=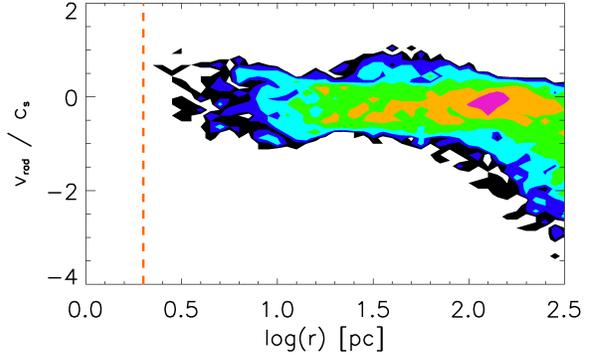,width=8.4cm}
\caption{The ratio of radial velocity to the sound speed of the gas, as a function of distance from the center of the halo.   As in Fig. 2, here contours denote the relative mass fraction of all of the gas shown, this fraction doubling across contour lines.  There is no significant supersonic inflow of gas (i.e. a 'cold flow') into the central $\sim$ 100 pc of the halo.  Indeed, the Mach number of the gas in the central region of the halo is $\le$ 2, unlike in the case where the primordial gas may cool in the absence of a photodissociating background radiation field (see e.g. Greif et al. 2008).  The dashed vertical line denotes the radius within which SPH particles can be 
accreted onto the sink particle representing the supermassive star expected to form at the center of the halo, as described in Section 3.  Note that negative velocities here correspond to infall.   
}
\end{figure}

For our simulation of the assembly of a primordial atomic-cooling DM halo at $z$ $\sim$ 15, we have employed multi-grid initial 
conditions which offer higher resolution in the region where the halo forms (e.g. Kawata \& Gibson 2003).
We initialize the simulation according to the $\Lambda$CDM power spectrum at $z$ = 100, adopting the cosmological 
parameters $\Omega_{m}=1 - \Omega_{\Lambda}=0.3$, $\Omega_{B}=0.045$, $h=0.7$, and $\sigma_{8}=0.9$, similar to the 
values measured by the {\it Wilkinson Microwave Anisotropy Probe} in its first year (Spergel et al. 2003). We use a 
periodic box with a comoving size $L$ = 1 $h^{-1}$ Mpc for the parent grid.  Our simulations use $N_{\rm DM}$ = $N_{\rm SPH}$ 
= 1.05 $\times$ 10$^6$ particles for DM and gas, where the SPH particle mass is $m_{\rm SPH}$ $\sim$  120 ${\rm M}_{\odot}$ in the region 
with the highest resolution. For further details on the technique employed to generate our multi-grid initial 
conditions, see Greif et al. (2008).  The maximum gas density that we resolve is $n_{\rm res}$ $\sim$ 10$^4$ cm$^{-3}$, 
higher than in previous work using similar initial conditions (Johnson et al. 2008, 2009) owing to the effect of the elevated 
LW background radiation field that we have included at a level of $J_{\rm LW}$ = 10$^3$ $\times$ 10$^{-21}$ erg s$^{-1}$ cm$^{-2}$ Hz$^{-1}$ sr$^{-1}$. 
Following Shang et al. (2010), we account for the shielding of the gas from this radiation by calculating the column density of H$_2$ 
over the local Jeans length and then applying the prescription for H$_2$ self-shielding provided by Draine \& Bertoldi (1996).  
As also described in Shang et al. (2010), we include the photodisscociation of H$^-$, assuming a background radiation temperature of 10$^4$ K. 
This is appropriate for a radiation field generated by standard Pop~II stars, which may constitute the bulk of the star formation already at $z$ $\la$ 15 
(see Maio et al. 2010).

\begin{figure*}
\includegraphics[width=7.0in]{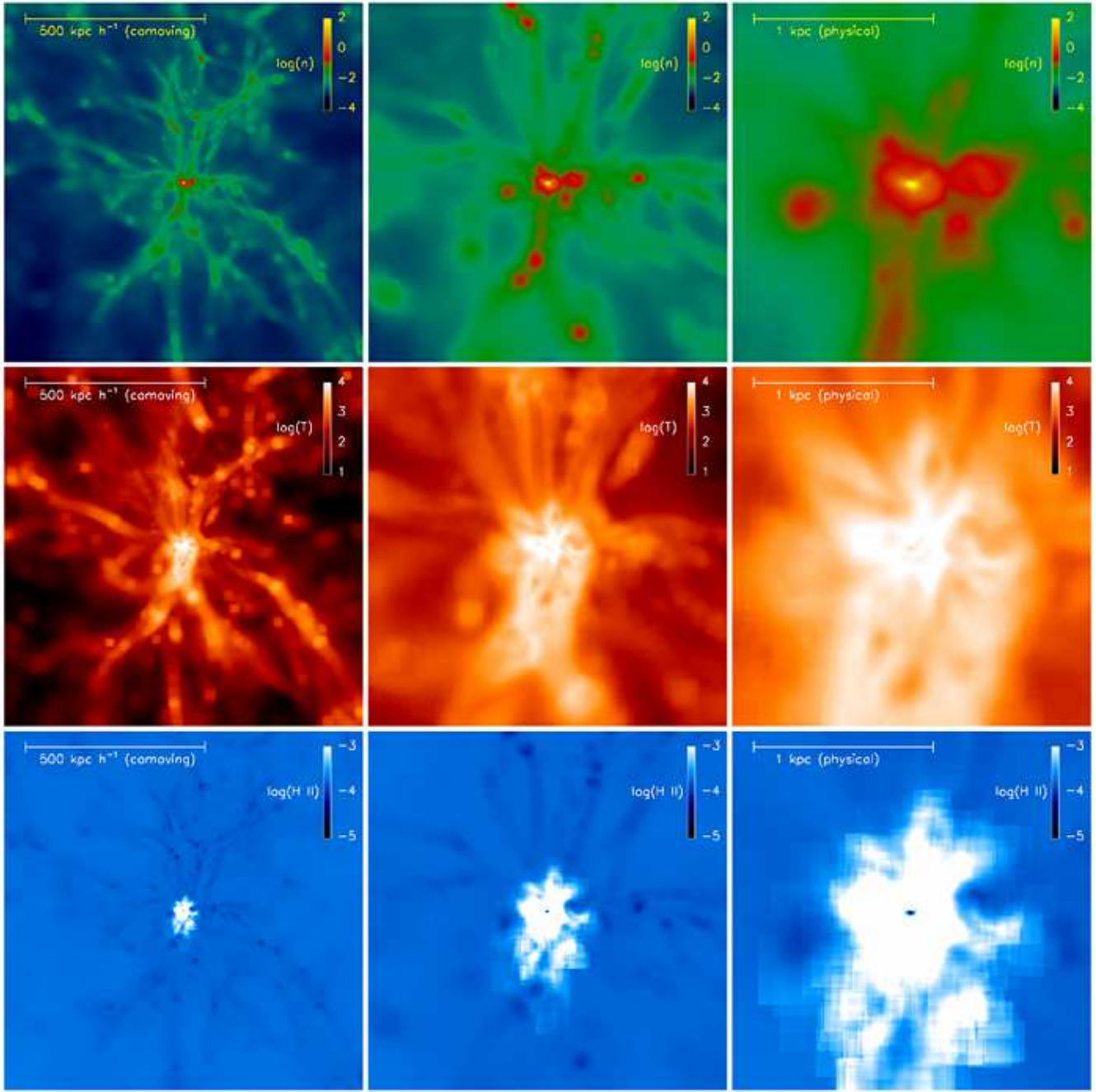}
\caption{The projected number density ({\it top row}), density-weighted temperature ({\it middle row}), and density-weighted H~{\sc ii} fraction ({\it bottom row}) of the gas in the simulated cosmological volume at $z$ $\sim$ 15, at which point a sink particle representing the central accreting object in the atomic-cooling halo is created at the location of the most dense SPH particle.  From left to right the panels show the properties of the gas at different scales, as labeled.  Note that the temperature of the gas within the $\sim$ 600 pc virial radius of the central halo can exceed 10$^4$ K, leading to collisional ionization of atomic hydrogen.}
\end{figure*}

\subsection{Properties of the Gas}
Figure 1 shows the properties of the primordial gas, as a function of the distance from the densest SPH particle in the most massive DM halo in our cosmological box
at $z$ $\sim$ 15.  At this point, $\la$ 10$^6$ M$_{\odot}$ of gas with a comparable Jeans mass has accumulated at the center 
of a halo with a total mass of $\sim$ 4 $\times$ 10$^7$ M$_{\odot}$ and a virial temperature of 1.8 $\times$ 10$^4$ K.  Thus, the conditions are in place for the collapse 
of the gas to form a massive central object which will likely accrete a sufficient amount of gas to form a supermassive star (SMS) and, shortly thereafter,
collapse directly to form a BH (e.g. Bromm \& Loeb 2003; see also Volonteri \& Begelman 2010).  

As shown in Figs. 1 and 2, the temperature of the gas in the densest regions of the halo spans a range, but is generally well above $\sim$ 10$^3$ K, owing to the low 
H$_2$ fraction of the gas, $f_{\rm H2}$ $\la$ 10$^{-6}$, which lead to cooling times of e.g. $\ga$ 10$^8$ yr at densities $\la$ 10 cm$^{-3}$ and temperature $\sim$ 5 $\times$ 10$^3$ K.  This is in agreement with previous simulations which have tracked the evolution of the gas collapsing into atomic-cooling haloes 
(Wise et al. 2008; Regan \& Haehnelt 2009b; Shang 2010).  The density profile of the gas is also in agreement with what these authors have found.
Furthermore, we find that the collapsing gas is only mildly turbulent in the center of the halo, with Mach numbers $\la$ 2, as has been found in 
similar simulations which neglected H$_2$ cooling (Wise \& Abel 2007).  However, as shown in Fig. 3, we do not find appreciable supersonic infall of 
gas, i.e. cold flows, owing to the inability of the primordial gas to cool in the central regions of the halo, unlike has been found in previous simulations 
in which H$_2$ cooling was not inhibited by a LW background (Greif et al. 2008).  As is illustrated in Fig. 4, 
a large portion of the gas within the virial radius, $\sim$ 600 pc from the center of the halo, has been shock heated to near the virial temperature of $\sim$ 1.8 $\times$ 10$^4$ K, leading to the partial collisional ionization of hydrogen and to the efficient cooling of the gas through atomic transitions.   

Figure 5 shows the angular momentum distribution of the gas in the central region of the halo, which 
we also find to be in rough agreement with previous work (e.g. Regan \& Haehnelt 2009b).  
The angular momentum of the gas has a critical impact on how much mass the central collapsed object, presumably a SMS, 
can accrete before it finally collapses to form a massive BH, as is discussed in the next Section.

\begin{figure}
\vspace{2pt}
\epsfig{file=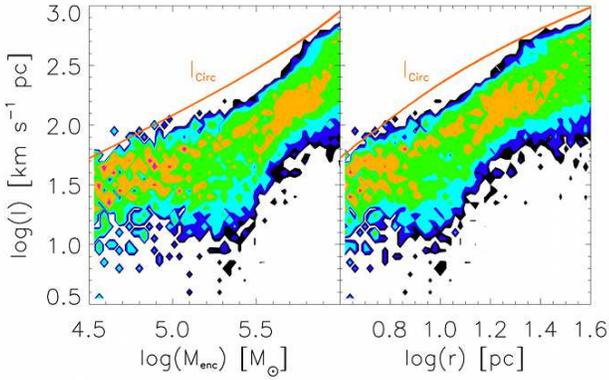,width=8.4cm}
\caption{The specific angular momentum of the gas about the most dense SPH particle at $z$ $\sim$ 15, as a function of distance ({\it right panel}) and of the enclosed mass ({\it left panel}).  As in Fig. 1, contours denote the distribution of the gas at a fixed radius $r$ and at fixed mass coordinate $M_{\rm enc}$, in the right and left panels, respectively.
 The specific angular momentum needed for circularization of the gas, defined as $l_{\rm Circ}$ = ($G$$M_{\rm enc}$$r$)$^{1/2}$, is shown by the orange lines.  At the radii resolved in our simulation, the gas is in general not rotationally supported and of the order of 10$^5$ M$_{\odot}$ of gas may be accreted onto the sink particle representing the collapsed massive object in the center of the halo.  However, the flattening of the $l$ profile at $r$ $\le$ 10 pc indicates that a rotationally supported disk is likely to form at radii that are not resolved in our simulation (see also Regan \& Haehnelt 2009b).  
}
\end{figure}

\begin{figure}
\vspace{2pt}
\epsfig{file=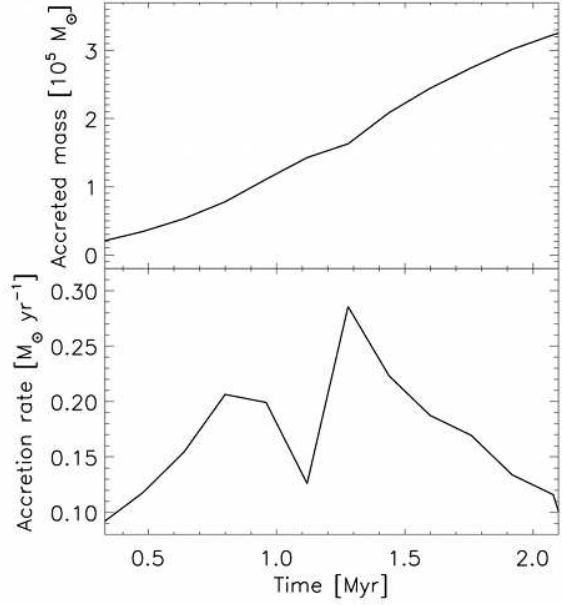,width=8.4cm}
\caption{The mass accreted by the sink particle representing the central object in the halo ({\it top panel}), presumed to be a SMS, and the mass accretion rate ({\it bottom panel}), as functions of time since the formation of the sink.  All SPH particles coming within 2 pc (the resolution length) of the sink on an orbit the semi-major axis of which is $\le$ 2 pc are accreted.  The sink grows to 3 $\times$ 10$^5$ M$_{\odot}$ in $\sim$ 2 Myr, the maximum lifetime for a SMS.  The high accretion rate, of the order of 0.1 M$_{\odot}$ yr$^{-1}$, may inhibit the escape of ionizing radiation during it's rapid growth. Some fraction of this accreted gas is presumed to end up in a massive black hole.      
}
\end{figure}

\section{evolution and accretion onto the central object}
Before the formation of a BH by direct collapse, 
we expect that the gas in the center of the halo first collapses to form a SMS, the immediate precursor to the BH.  In this Section, we
study the accretion of gas onto this SMS in the absence of radiative feedback.

Beginning from the point of the simulation at $z$ $\sim$ 15, the state of which is shown in Figures 1-5, we calculate 
the rate at which gas accretes onto the SMS.  
To this end, we employ the same sink particle routine as used in Johnson et al. (2008), and we adopt a simple prescription for 
the accretion of SPH particles onto the sink particle, which represents the unresolved region surrounding the SMS in our simulation
and is created at the location of the most dense SPH particle with an initial mass of $\sim$ 10$^4$ M$_{\odot}$, roughly equal to the mass in the kernel.  
Defined as the density at which the local Jeans mass becomes comparable to the SPH kernel mass, the maximum density that is resolved is $n_{\rm res}$ $\sim$ 
10$^4$ cm$^{-3}$; in turn, the minimum resolved length scale is roughly ($m_{\rm SPH}$ / ($m_{\rm H}$ $n_{\rm res}$))$^{\frac{1}{3}}$ $\sim$ 2 pc, where $m_{\rm H}$ is the mass of the hydrogen atom.  Therefore, conservatively, we allow SPH particles to be accreted onto the sink particle if they come within 2 pc of the sink and if the semimajor axis of their orbit about 
the sink is $\la$ 2 pc.  This allows us to resolve the evolution of 
the gas surrounding the sink particle and to obtain an upper limit for the mass of the SMS as it grows in time.  
We track the accretion of SPH particles for 2.2 Myr, roughly equal to  the $\sim$ 2 Myr maximum lifetime of a SMS (see e.g. Begelman 2010).

Figure 6 shows the rate at which mass accretes onto the sink particle and the total mass accreted, as functions of the time since
the formation of the sink.  
The accretion proceeds relatively smoothly, with variations in the accretion rate reflecting variations in the density of the accreting gas, shown in Figs. 1 and 4. 
The mass of the sink particle increases continuously at a rate 
0.1 M$_{\odot}$ yr$^{-1}$ $\la$  $\dot{M_{\rm acc}}$ $\la$ 0.3 M$_{\odot}$ yr$^{-1}$.  The total mass of the sink particle 
after 2 Myr is $\sim$ 3 $\times$ 10$^5$ M$_{\odot}$, consistent with the fraction, $\sim$ 2/3, of the the inner 10$^6$ M$_{\odot}$ 
of gas which was moving on infalling trajectories at speeds approaching the sound speed, just before the creation of 
the sink particle (see Fig. 3).  

It is important to note that the accretion rates that we find here are upper limits, for two reasons.  Firstly, the sink particle that 
is created exerts no pressure on the surrounding gas, which may lead to somewhat artificially enhanced infall velocities for the neighboring SPH particles.   
Secondly, at this stage we neglect any radiative feedback on the gas, which could act to inhibit accretion.
How important radiative feedback is in slowing the accretion of gas onto a SMS is not 
entirely clear.  However, at the high accretion rates that we find 
the infalling matter may be optically thick, trapping the ionizing radiation emitted by the SMS 
and so limiting the radiative feedback at the scales we resolve.   

Some fraction of the gas accreted onto the SMS is expected to end up in the BH that forms at its center; this fraction  
may be up to $\sim$ 90 percent (see Shibata \& Shapiro 2002), but may be perhaps an order of magnitude lower due to the copious 
high-energy radiation emitted from the BH as it grows in the center of the star, as discussed in  Begelman (2010).  Therefore, we 
shall consider that a range of initial BH masses, from $\sim$ 10$^4$ to $\sim$ 10$^5$ M$_{\odot}$, is possible.
In the next Sections, we study the accretion of gas onto the black hole that forms, for different choices of its initial mass.

\begin{figure}
\vspace{2pt}
\epsfig{file=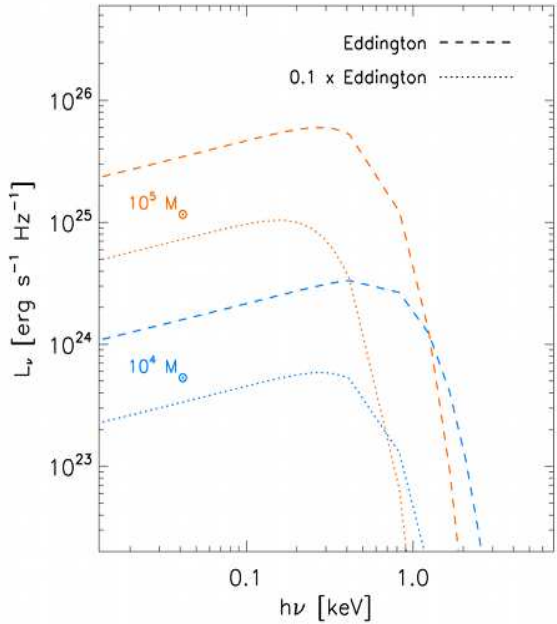,width=8.4cm,height=8.8cm}
\caption{Spectra of thermal radiation emitted by a multi-color BH accretion disk, for two illustrative black hole masses and accretion rates, in units of the Eddington accretion rate.  At every timestep in our simulations the spectra of the accretion disk is calculated as a function of the mass of the BH and the accretion rate.  From the spectrum are obtained the numbers of H~{\sc i}- and He~{\sc ii}-ionizing photons, the H$_2$ photodissociation rate and photoheating rates, and the radiation pressure on the surrounding gas due to Thomson scattering and photoionization.  
}
\end{figure}

\section{Modelling black hole accretion and radiative feedback}
In this Section, we describe our methodology for calculating the accretion rate onto BHs, 
as well as our treatment of the radiative output from the accretion disk, which is 
coupled in detail to the surrounding gas.

\subsection{The Accretion Rate}
The Bondi radius, which must be resolved in a simulation in order for the Bondi accretion rate onto a BH to be properly estimated, is given by

\begin{equation}
 R_{\rm Bondi} \sim 0.3 {\rm pc} \left(\frac{M_{\rm{BH}}}{10^4 {\rm M_{\odot}}}\right) \left(\frac{T}{10^4 {\rm K}}\right)^{-1} \mbox{\ ,}
\end{equation}
where $M_{\rm BH}$ is the mass of the BH and $T$ is the temperature of the gas surrounding the BH.  
We compare this to the smallest scale that can be resolved in our 
simulation; specifically, we compute the minimum length scale 
$R_{\rm res}$ of the SPH kernel for which the Jeans mass is larger than the mass of the kernel ($\sim$ 50 $M_{\rm SPH}$), given by

\begin{equation}
R_{\rm res} \sim 0.2 {\rm pc} \left(\frac{M_{\rm SPH}}{120 {\rm M_{\odot}}}\right)^{\frac{2}{3}} \left(\frac{T}{10^4 {\rm K}}\right)^{-1} \mbox{\ ,}
\end{equation}  
where $M_{\rm SPH}$ is the mass of an SPH particle; in our case, $M_{\rm SPH}$ = 120 M$_{\odot}$.  Thus, for the cases that we consider, 
with $M_{\rm BH}$ $\ga$ 10$^4$ M$_{\odot}$,
the Bondi radius will always be resolved and we can have confidence in the accretion rates that we find, at least at the scales we resolve
at which the gas is not rotationally supported, as shown in Fig. 3 (see Hopkins \& Quataert 2010; also Booth \& Schaye 2009).  
However, it should be noted that at smaller scales some fraction of the gas
may instead go into an accretion disk wind or otherwise be accreted on longer timescales than we assume here, 
such as over the viscous timescale of the accretion disk (see e.g. Power et al. 2010).  In this sense, the accretion rates 
that we find, calculated as described below, are likely to be upper limits.    

The accretion rate onto the BH is calculated directly as the Bondi (1952) accretion rate:

\begin{equation}
\dot{M}_{\rm{BH}} \sim \frac{4 \pi \left(G M_{\rm{BH}}\right)^2 \rho_{\rm{gas}}}{\left(c_{\rm s}^2 + v_{\rm BH}^2 \right)^{\frac{3}{2}}} \mbox{\ ,}
\end{equation}
where the gas sound speed $c_{\rm s}$ and gas density $\rho_{\rm gas}$ are taken as the average of these values over 
all SPH particles within the smoothing length of the sink particle which marks the position of the BH; 
finally, $v_{\rm BH}^2$ is the square of the velocity of the sink particle
with respect to the SPH particles within its smoothing length. 

We track the total accreted mass onto the BH and every time a mass equal to the mass of an 
SPH particle (120 M$_{\odot}$) is accreted, 
a random SPH particle lying within a smoothing length of the BH 
is accreted, using the same sink particle routine as discussed in the previous Section.

\subsection{Radiation Emitted from the Accretion Disk}
At every timestep,  we compute the spectrum of the radiation emitted from the accretion disk of the BH, 
as a function of the mass of the BH and the accretion rate, as follows.  The temperature $T$ of the accretion disk, as a function of distance $r$ from the BH, is given by

\begin{equation}
T(r)=\left(\frac{3}{8\pi}\frac{GM_{\rm{BH}}\dot{M}_{\rm{BH}}}{\sigma_{\rm{SB}} r^{3}}\right)^{\frac{1}{4}}\mbox{\ ,}
\end{equation}
where $M_{\rm{BH}}$ is the mass of the BH, $\dot{M}_{\rm{BH}}$ the accretion rate, and $\sigma_{\rm{SB}}$ the Stefan-Boltzmann constant (Pringle 1981). 
For simplicity, we have taken the disk to be a thin disk, such that each annulus radiates as a blackbody of temperature given by the above equation.  
Integrating the flux over the surface of the disk, from $r_{\rm inner}$ to $r_{\rm outer}$ = 10$^4$ $r_{\rm inner}$, 
at which distance the contributions to both the photodissociating and ionizing fluxes are negligible, we find the total emitted flux as a function of the BH mass and its accretion rate.
We take for the inner-most radius of the disk 

\begin{equation}
r_{\rm inner} \sim 2 {\rm km} \left(\frac{M_{\rm{BH}}}{{\rm M_{\odot}}}\right) \mbox{\ , } 
\end{equation}
corresponding to a high value for the black hole spin parameter $a$ $\ga$ 0.9 (e.g. Makishima et al. 2000; Vierdayanti et al. 2008), 
which we expect considering that at its formation the BH accretes stellar material which can have a large angular momentum (see e.g. Volonteri \& Begelman 2010).

Figure 7 provides several examples of the spectra that we obtain using this treatment, for different choices of the BH mass and accretion rate.  
For a given BH mass, a higher accretion rate results in a higher bolometric luminosity and also in emission at higher energies.  However,
for accretion of gas at a given fraction of the Eddington rate, the peak of the spectrum is at higher energy for a less massive BH, owing to the smaller inner
radius of the accretion disk at which the temperature is higher.

Recent observations of the X-ray emission from an accreting BH with a mass 10$^2$ $\la$ $M_{\rm BH}$ $\la$ 10$^5$ M$_{\odot}$ (Farrell et al. 2010), 
bracketing the range of BH masses that we consider in the present study,
suggest that the bulk of the emission at energies $\la$ 1 keV can be accounted for by the presence of an accretion disk radiating as a blackbody, as we assume here, 
while above this energy
a power law component is shown to provide an appropriate fit to the observed spectrum.  For simplicity, we do not include a power law component 
in the spectrum emitted from the accreting BH.  Including one, however, would not likely make a significant difference in the strength of the radiative 
feedback on the gas within the host halo, as the optical depth $\tau$($\nu$) for photons with energy $h$$\nu$ is 

\begin{equation}
\tau(\nu) \sim \sigma(\nu) \Omega_{\rm b} M_{\rm halo} / (4 \pi r_{\rm vir}^2 m_{\rm H}) \sim 10^{-2}  \left(h \nu / 1 {\rm keV}\right)^{-3} \mbox{\ , }
\end{equation}
where $\sigma$($\nu$) $\sim$ 4 $\times$ 10$^{-20}$ cm$^2$ ($h$$\nu$/0.1 keV)$^{-3}$ is the photoionization cross section for neutral primordial 
gas (e.g. Kuhlen \& Madau 2005), $\Omega_{\rm b}$ is the cosmological baryon fraction, $M_{\rm halo}$ $\sim$ 3.5 $\times$ 10$^7$ M$_{\odot}$ is the 
total mass of the halo, $r_{\rm vir}$ $\sim$ 600 pc is the virial radius of the halo, $m_{\rm H}$ is the mass of the hydrogen atom. At $h$$\nu$ 
$\ga$ 1 keV, $\tau$ $\la$ 10$^{-2}$; hence, most of these photons would not interact with the gas in the halo and would contribute little to 
the photoionization of the gas.  We point out, though, that this does have implications for our estimate of the escape fraction of ionizing photons, as discussed in Section 5.4.

\subsection{Radiative Feedback on the Gas}
To model the propagation of the high energy radiation emitted from the accretion disk, we use a modified version 
of the ray-tracing routine described in Johnson et al. (2009) and Greif et al. (2009).  We refer the interested reader to those works for a detailed description
of the scheme that we use to treat the propagation of ionizing radiation by solving, at every timestep, for the H~{\sc ii} and H~{\sc iii} regions surrounding the accreting BH;
here we will briefly describe the improvements and modifications that have been made to that scheme. 

The main modification in the calculation of the photoheating and photoionization rates is that here we 
adopt a multi-color disk spectrum appropriate for a black hole accretion disk, as described in Section 4.2, instead of a stellar spectra as in previous work.
We integrate over the accretion disk radius $r$ and over the frequency of the thermal radiation $\nu$ emitted from the disk to calculate the photoionization 
and associated photoheating rates for both the photoionization of both H~{\sc i} and He~{\sc ii}, as well as to obtain the total numbers of ionizing photons, 
$Q_{\rm H II}$ and $Q_{\rm He III}$, emitted per second from the disk for each of these reactions, respectively.

As well, we estimate the photodissociation rate of H$_{\rm 2}$, by finding the total flux emitted from the accretion disk at $12.87~\rm{eV}$ (e.g. Abel et al. 1997).  
We account for the self-shielding of H$_2$ in our ray-tracing routine by calculating the column density of molecules along each ray and using the prescription 
provided by Draine \& Bertoldi (1996); however, we find that 
the column densities realized in our simulations are too low for self-shielding to have an effect in lowering the photodissociation rate. 

Finally, as follows, we account for the transfer of momentum to the gas due to the anisotropic radiation pressure from both electron scattering and photoionization, 
which can have an important effect on the dynamics of the primordial gas (Milosavljevi{\'c} 2009a,b; Park \& Ricotti 2010; see also Haehnelt 1995).
The radiation pressure on the $i$th SPH particle due to electron scattering is estimated as

\begin{equation}
P_{\rm elec, i} \sim \frac{L_{\rm acc} \sigma_{\rm T} n_{\rm elec, i} V_{\rm i}^{\frac{1}{3}}}{4 \pi r_{\rm i}^2 c} \mbox{\ ,}
\end{equation}
where $L_{\rm acc}$ is the total luminosity emitted from the accretion disk, $n_{\rm elec, i}$ is the electron number density of the gas represented by the SPH 
particle, $r_{\rm i}$ is the distance from the BH to the particle, $c$ is the speed of light, and $\sigma_{\rm T}$ is the Thomson cross section.  
The approximate volume $V_{\rm i}$ of gas represented by the particle is estimated as $V_{\rm i}$ $\sim$ $M_{\rm i} / \rho_{\rm i}$, 
where $M_{\rm i}$ is the mass of the particle and $\rho_{\rm i}$ is its density.  The radiation pressure due to photoionization is estimated as

\begin{equation}
P_{\rm ion, i} \sim \frac{<E_{\rm \gamma}> \left(k_{\rm HI} n_{\rm HI, i} + k_{\rm HeII} n_{\rm HeII, i}\right) V_{\rm i}^{\frac{1}{3}}}{c} \mbox{\ ,}
\end{equation} 
where $<E_{\rm \gamma}>$ is the average energy of ionizing photons emitted from the accretion disk, 
$n_{\rm HI, i}$ and $n_{\rm HeII, i}$ are the number densities of H~{\sc i} and He~{\sc ii}, and  
$k_{\rm HI}$ and $k_{\rm HeII}$ are the photoionization rates of H~{\sc i} and He~{\sc ii}. 
These two pressures added together yield the total radiation pressure on particle $i$, $P_{\rm rad, i}$ = $P_{\rm elec, i}$ + $P_{\rm ion, i}$.  In general, we find 
that the photoionization pressure is much greater than the electron scattering pressure, due to the non-negligible fractions of H~{\sc i} and He~{\sc ii} in photoionized regions.

\begin{figure*}
\includegraphics[width=7.0in]{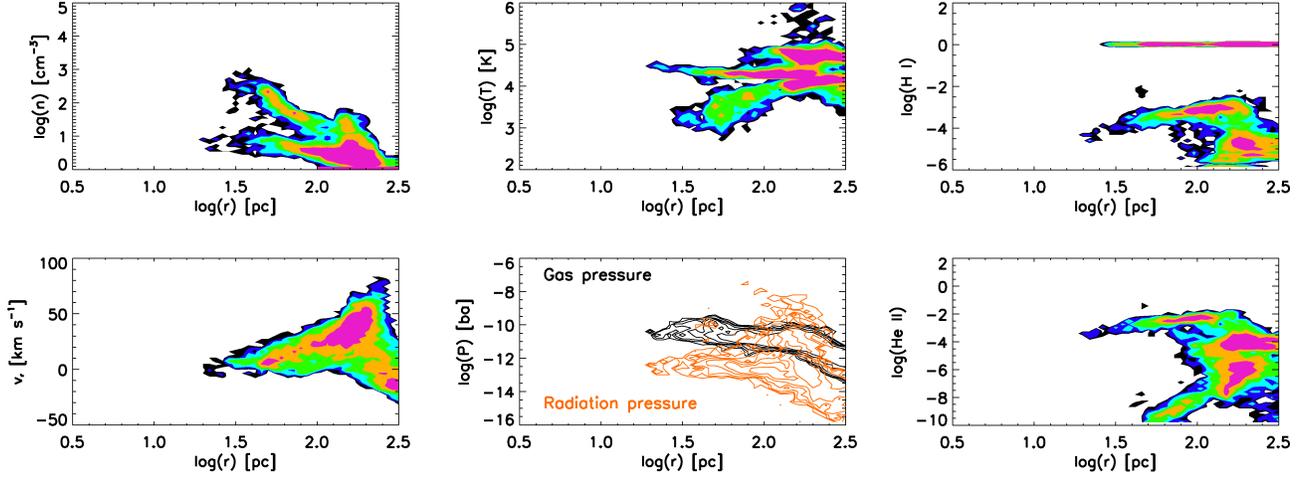}
\caption{Properties of the gas at the end of the simulation, 3 Myr after the formation of the BH, as a function of physical distance from the BH, for the case of a 5 $\times$ 10$^4$ M$_{\odot}$ initial BH mass, with the contours denoting the relative mass fraction of all of the gas shown.  Clockwise from top-left: number density of hydrogen nuclei, temperature, neutral hydrogen (H~{\sc i}) fraction, singly ionized helium (He~{\sc ii}) fraction, gas and radiation pressure, and radial velocity.  Strong photoheating and photoionization pressure drive the gas out from the center of the halo at up to $\sim$ 80 km s$^{-1}$, although some relatively cold neutral gas remains in the vicinity of the BH, as it is shielded from the photoionizing radiation.  While the radiation pressure is higher than the gas pressure where the neutral fraction is highest (see eq. 8), in general the gas pressure is dominant.  In the photoionized region at $r$ $\ge$ 100 pc from the BH, the primary coolant is He~{\sc ii} instead of H~{\sc i}, yielding higher temperatures in this region than in the higher density central region of the halo.}
\end{figure*}

\begin{figure*}
\includegraphics[width=7.0in]{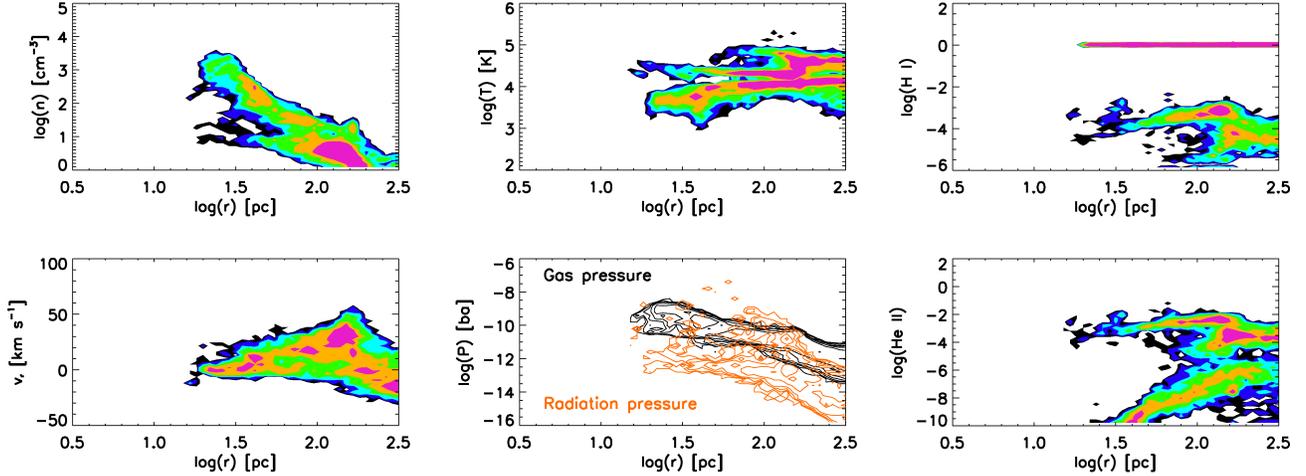}
\caption{Just as Fig. 8, but for the case of a 2.5 $\times$ 10$^4$ M$_{\odot}$ initial BH mass.  The gas properties are similar to those found in the 5 $\times$ 10$^4$ M$_{\odot}$ case, although the temperature of the photoionized gas at large radii is somewhat lower in this case, as is the speed at which the gas is pushed outward from the center of the halo, largely owing to the initially much higher accretion rate and ionizing photon output of the more massive BH (see Figs. 12 and 13).
}
\end{figure*}

\begin{figure*}
\includegraphics[width=7.0in]{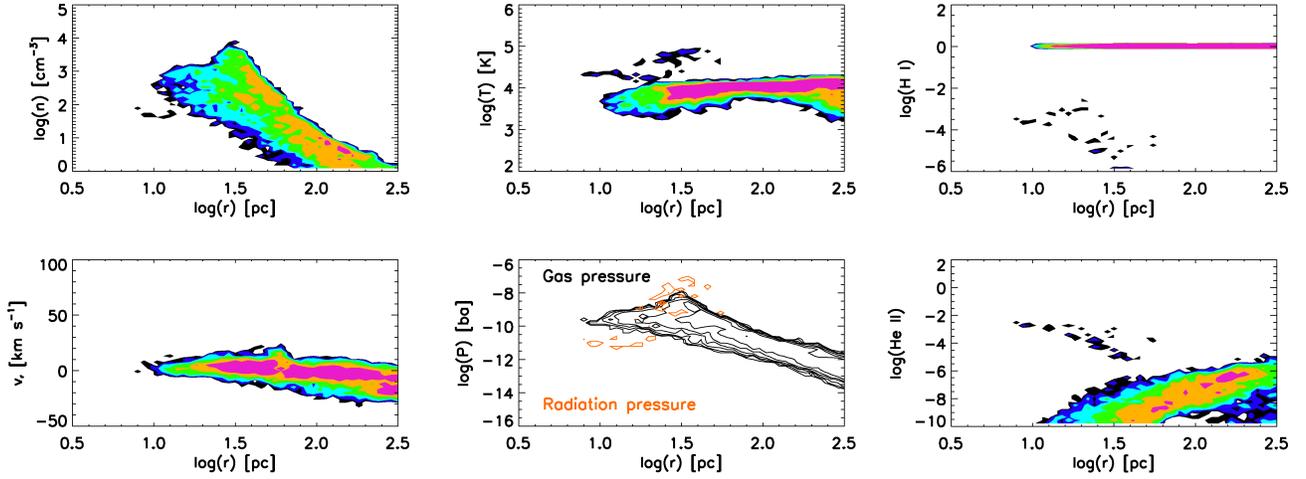}
\caption{Just as Fig. 8, but for the case of a 10$^4$ M$_{\odot}$ initial BH mass.  Only a small fraction of the gas near the BH are ionized in this case, owing to the lower number of ionizing photons emitted from the accretion disk (see Fig. 13).  As a result, the density structure of the gas closest to the BH is less disrupted than in the cases of the more massive BHs (see also Fig. 14).
}
\end{figure*}

We then calculate the acceleration, in the direction opposite that of the BH, of SPH particles lying within the photoionized region as

\begin{equation}
a_{\rm rad, i} \sim \frac{P_{\rm rad, i} V_{\rm i}^{\frac{2}{3}}}{M_{\rm i}} \mbox{\ ,}
\end{equation}
which is expressed more precisely as
\begin{equation}
a_{\rm rad, i} = \left(m_{\rm H} c \right)^{-1}\left[\frac{L_{\rm acc} \sigma_{\rm T} f_{\rm elec, i}}{4 \pi r_{\rm i}^2} + <E_{\rm \gamma}> \left(k_{\rm HI} f_{\rm HI, i} + k_{\rm HeII} f_{\rm HeII, i}\right)\right] \mbox{\ ,}
\end{equation} 
where $f_{\rm elec,i}$, $f_{\rm HI,i}$, and $f_{\rm HeII,i}$ are the free electron fraction, H~{\sc i} fraction, and He~{\sc ii} fraction, respectively.
Finally, the effect of the radiation pressure is realized by imparting a velocity kick of magnitude $\delta$$v_{\rm rad, i}$ to the SPH particle in the direction opposite the BH:  

\begin{equation}
\delta v_{\rm rad, i} = a_{\rm rad, i} \delta t \mbox{\ ,}
\end{equation}
where $\delta$$t$ is the length of the timestep taken in the simulation, over which the radiative pressure is applied.  

We emphasize that in this treatment we have assumed that the radiation is emitted isotropically from the BH accretion disk, and in this 
we have not accounted for the absorption of ionizing radiation in the outer regions of the disk.  As this could effectively shield a 
substantial portion of the surrounding gas from the high energy radiation, the overall 
strength of the radiative feedback on the gas in our simulations is likely to be an overestimate.

\section{Results on black hole accretion and radiative output}
In this Section, we present the results of our simulations of BH accretion and the associated radiative feedback.  
The evolution of the gas within the host halo, the accretion rate of gas onto the BH, the radiative output from the accretion disk, 
and the escape of ionizing radiation from the host halo are all closely interconnected.  We discuss each of these, in turn.

Using as initial conditions the output from the cosmological simulations described in Section 2 after the creation of the central sink particle, which we
take to represent the accreting BH, we ran three separate simulations, each with a different initial BH mass.  These initial BH masses are 
$M_{\rm BH, init}$ = 10$^4$, 2.5 $\times$ 10$^4$, and 5 $\times$ 10$^4$ M$_{\odot}$.

\subsection{Evolution of the Gas}
With the start of BH accretion, the radiative output from the accretion disk dramatically impacts the state of the surrounding gas.  
Figures 8, 9, and 10 show the properties of the gas, for the simulations with initial BH masses of 5 $\times$ 10$^4$, 2.5 $\times$ 10$^4$, and 10$^4$ M$_{\odot}$, respectively, 
after 3 Myr since the formation of the BH.  Comparing each of these Figures, it is clear that accretion onto the more massive BHs leads to stronger radiative feedback on the gas.
To illustrate the main effects, we shall thus focus on the case of the most massive BH, shown in Fig. 8.

As shown in the top panels in Fig. 8, there is a two-phase medium that develops within $\sim$ 100 pc of the BH, 
with the gas that is photoionized having, in general, lower densities ($\la$ 10 cm$^{-3}$) and higher temperatures ($\ga$ 10$^4$ K), 
while the more dense gas that is shielded from the ionizing radiation remains at $\la$ 10$^4$ K.  
The photoheated gas experiences a boost in both gas pressure and radiation pressure, 
which results in the gas being pushed outward at speeds up to $v_{\rm r}$ $\sim$ 80 km s$^{-1}$, as shown in the bottom-left panel.
 
\begin{figure}
\vspace{2pt}
\epsfig{file=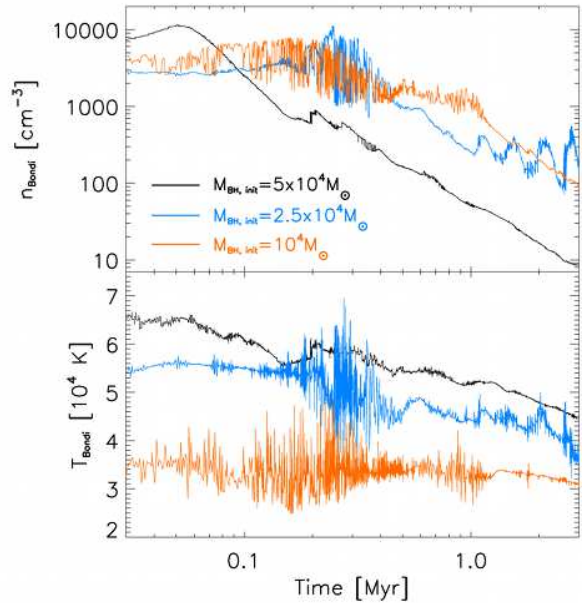,width=8.4cm}
\caption{The density ({\it top panel}) and the temperature ({\it bottom panel}) in the vicinity of the BH that are used to compute the Bondi accretion rate, as functions of time, for the three different cases of BH mass in our study.  The temperature varies much less than the density over the 3 Myr of our simulations; hence, it is largely variations in the density of the accreted gas that affect changes in the accretion rate (see Fig. 12) and in the strength of the associated radiative feedback (see Fig. 13).
}
\end{figure}

As shown in the bottom-middle panel of Fig. 8, the gas pressure is generally much higher than the radiation pressure, execpt in photoionized regions where the fractions of H~{\sc i} and He~{\sc ii} are relatively high,
as it is here that the photoionization rates per unit volume are highest (see eq. 8).  In particular, this occurs at $\sim$ 100-200 pc from the BH, as shown in the panels on the right, where 
the radiation pressure therefore plays an important role in driving the gas out from the center of the halo.
Interestingly, at distances $\ga$ 100 pc from the BH we find that the temperature of the ionized gas jumps to $\la$ 10$^5$ K; this occurs where the He~{\sc ii} fraction is highest, leading 
to an enhanced photoheating rate through the reaction He~{\sc ii} + $\gamma$ $\to$ He~{\sc iii} + e$^-$.  This increased heating rate leads to temperatures high 
enough that hydrogen is almost completely collisionally ionized, leading in turn to a decreased cooling rate by atomic hydrogen; in effect, the gas approaches thermal equilibrium with He~{\sc ii} as 
the primary coolant at $r$ $\ga$ 100 pc, whereas H~{\sc i} is the primary coolant closer to the BH where the gas density is higher.  
A similar effect has been reported by Meiksin et al. (2010). 


\begin{figure}
\vspace{2pt}
\epsfig{file=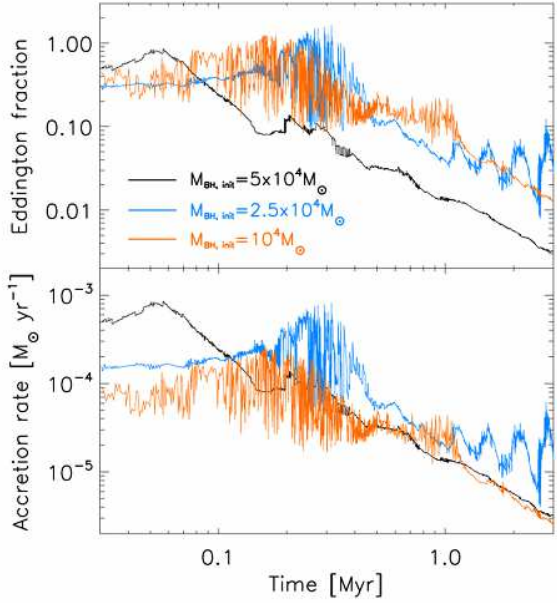,width=8.4cm}
\caption{The BH accretion rate ({\it bottom panel}) and Eddington ratio ({\it top panel}), as functions of time since the formation of the BH, for the three different BH masses in our study.  The accretion rates generally decrease with time, largely due to the expansion of the gas in response to photoheating and radiation pressure (see Figs. 8 - 11); in each of the cases, the Eddington fraction becomes roughly unity before dropping by at least an order of magnitude over the 3 Myr timescale of the simulations.           
}
\end{figure}

\subsection{Accretion Rate and Eddington Fraction}
The accretion rate, as described in Section 4.1, is directly coupled to the properties of the gas in the vicinity of the BH.  
Figure 11 shows the evolution of the temperature $T_{\rm Bondi}$ = 3$m_{\rm H}$$c_{\rm s}$$^2$/k$_{\rm B}$ and density $n_{\rm Bondi}$ = $\rho_{\rm gas}$/$m_{\rm H}$ 
in the vicinity of the BH that are used to compute the accretion rate $\dot{M}_{\rm{BH}}$ (see equ. 3), as a function of the time since the formation of the BH.  Comparing the top and bottom panels, it is clear that the
temperature of the accreted gas varies, at most, by a factor of $\la$ 2, whereas the density of the gas decreases by at least an order of magnitude in all cases, owing to the 
radiative feedback on the gas.  Therefore, we conclude that it is largely the evolution of the density which dictates the evolution of the accretion rate.

Indeed, as shown in Fig. 12, at the outset of the simulations the accretion rates are higher for the more massive BHs, and the Eddington fractions are comparable in all three cases.
However, as the hydrodynamic response of the gas to the radiative feedback develops and the density drops, 
the accretion rates drop dramatically.  The accretion rate of the 5 $\times$ 10$^4$ M$_{\odot}$ BH plummets by two orders of magnitude
over the 3 Myr timescale of the simulation.  The hydrodynamical response is not as swift in the other two cases, and the 10$^4$ and the 2.5 $\times$ 10$^4$ M$_{\odot}$ BHs experience 
brief periods of accretion at near the Eddington rate before the radiative feedback on the gas drives the gas density down, ultimately resulting in the accretion rate dropping by an order of 
magnitude.  For the majority of the time, all three BHs accrete at well below 10 percent of the Eddington rate.

\begin{figure}
\vspace{2pt}
\epsfig{file=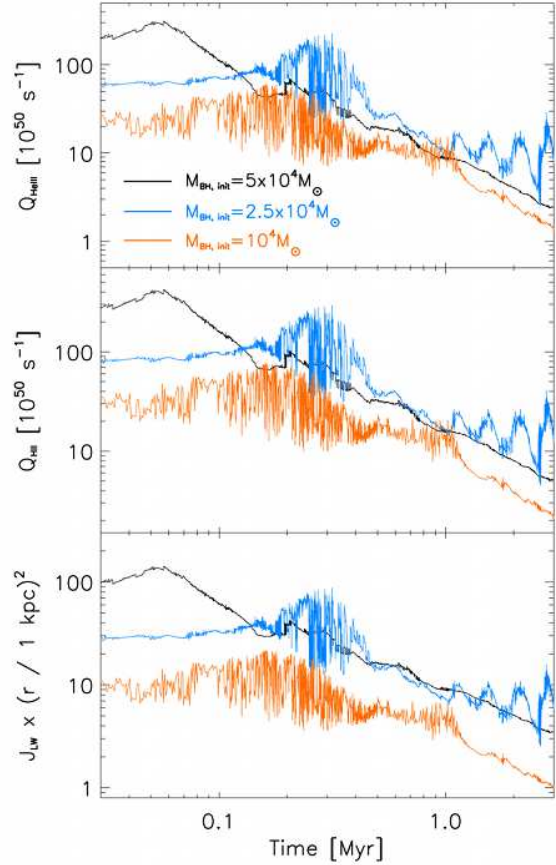,width=8.4cm}
\caption{
Properties of the high energy radiation emitted from the BH accretion disk, as functions of time since the formation of the BH, for each of the three 
BH masses in our study: the number of He~{\sc ii}-ionizing photons emitted per second ({\it top panel}), the number of 
H~{\sc i}-ionizing photons emitted per second ({\it middle panel}), and the molecule photodissociating flux $J_{\rm LW}$ 
(in units of 10$^{-21}$ erg s$^{-1}$ cm$^{-2}$ Hz$^{-1}$ sr$^{-1}$), 
normalized to a physical distance of 1 kpc from the BH ({\it bottom panel}).  The flux of ionizing photons in all three cases is high enough to substantially heat and ionize the gas in the vicinity of the BH.  While the flux of high energy radiation is initially higher for the more massive BHs, at later times this trend is reversed, reflecting the complex interplay between BH accretion
and the resulting radiative feedback on the gas.
}
\end{figure}

\begin{figure*}
\includegraphics[width=7.0in]{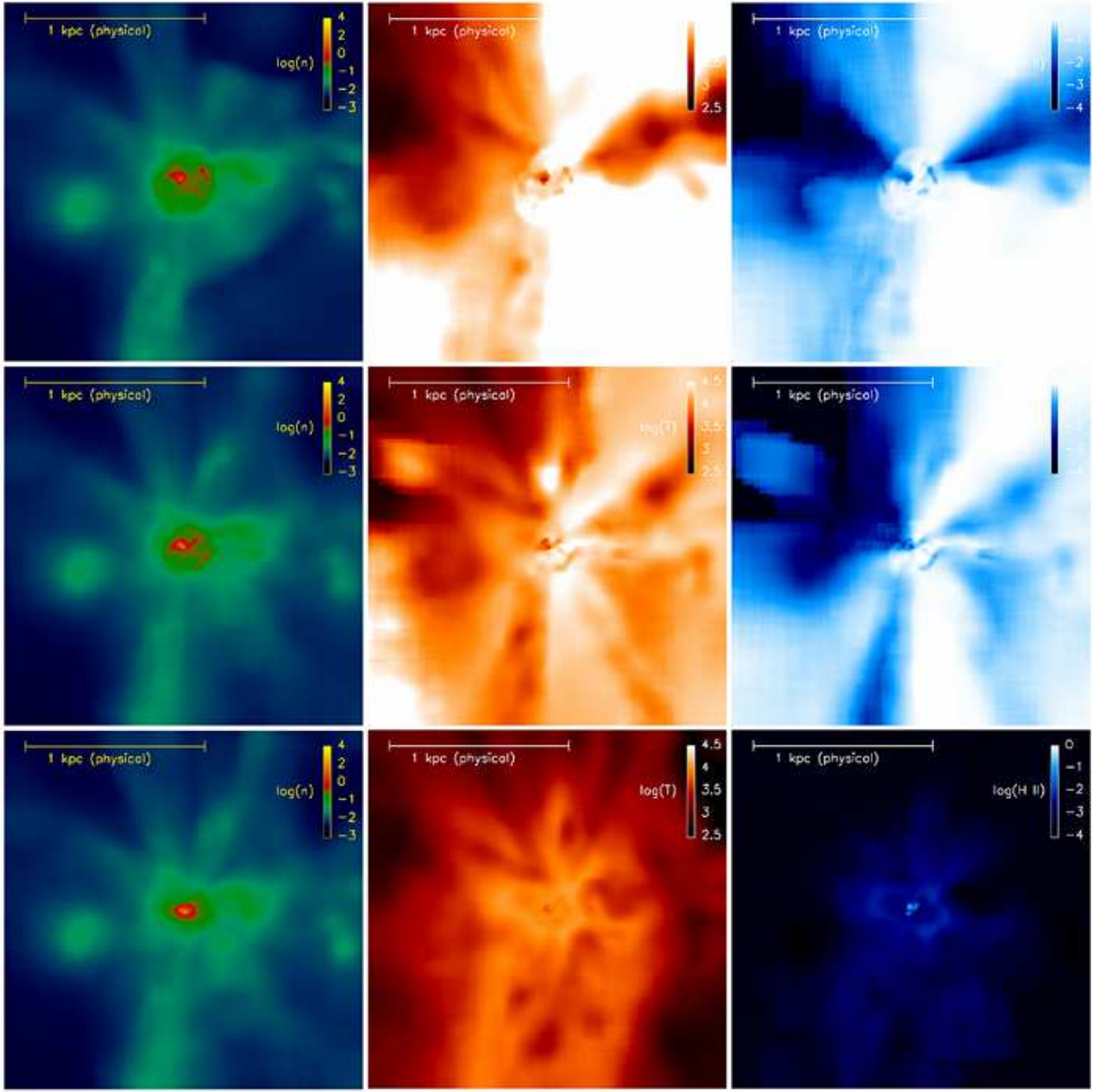}
\caption{The projected number density ({\it left column}), density-weighted temperature ({\it middle column}), and density-weighted H~{\sc ii} fraction ({\it right column}) of the gas in the vicinity of the accreting BH after 3 Myr since the formation of the BH, for the three cases of BH mass that we consider:  10$^4$ M$_{\odot}$ ({\it bottom row}), 2.5 $\times$ 10$^4$ M$_{\odot}$ ({\it middle row}), and 5 $\times$ 10$^4$ M$_{\odot}$ ({\it top row}).  The radiation emitted from the accretion disks of the more massive BHs has a stronger impact on the gas in the halo, leading to generally higher temperatures and to expansion of the gas causing expansion and disruption of the dense gas in the center of the halo from which the BH feeds (see also Figs. 8, 11-13).  Note also that more ionizing radiation escapes the halo for the cases of the more massive BHs (see Fig. 15).}
\end{figure*}

\subsection{Ionizing and H$_2$-Dissociating Photon Output}
We integrate over the spectrum emitted from the accretion disk to compute the 
total number of ionizing photons emitted per unit time.  The numbers of H~{\sc i}-
and He~{\sc ii}-ionizing photons emitted per unit time, $Q_{\rm HII}$ and $Q_{\rm HeIII}$, are defined here to be the total number of photons emitted
with energies above 13.6 eV and 54.4 eV, the ionization thresholds of H~{\sc i} and He~{\sc ii}, 
respectively.  The numbers of ionizing photons emitted per second are shown in the top two 
panels of Figure 13.  Comparing Figs. 12 and 13, it is clear that the ionizing photon 
production rate is strongly correlated with the accretion rate.  Indeed, we see the same 
trends in the production rate of ionizing radiation as for the accretion rate, with the 
more massive BHs initially emitting photons at a higher rate and the overall emission rates 
dropping by at least an order of magnitude for all three BH masses.

Due to the hard spectra emitted from the accretion disks (see Fig. 7), for all three initial BH masses,
the production rate of He~{\sc ii}-ionizing photons is almost as high as that of H~{\sc i}-ionizing photons, 
as shown in the top two panels of Fig. 13.
Indeed, the fraction $Q_{\rm HeIII}$/$Q_{\rm HII}$ is considerably higher than is expected for even 
very massive Pop III stars (see e.g. Bromm et al. 2001; Schaerer 2002).
As we discuss further in Section 6, this translates into a distinctive observational signature of 
the accretion of primordial gas onto BHs in the Early Universe.

In the bottom panel of Fig. 13 is shown the flux of molecule-dissociating (Lyman-Werner) radiation $J_{\rm LW}$, 
in the standard units of 10$^{-21}$ erg s$^{-1}$ cm$^{-2}$ Hz$^{-1}$ sr$^{-1}$, normalized to a distance $r$ = 1 physical kpc from the BH.
Interestingly, the photodissociating flux is almost always lower than the background flux assumed in our simulation ($J_{\rm LW}$=10$^3$) by 
at least an order of magnitude, at the fiducial distance of 1 kpc.  Therefore, the radiative output from the BH significantly impacts only 
the gas at much smaller distances from the BH.  However, the flux near the center of the halo, where the most dense gas resides, is easily 
high enough to suppress the cooling of gas by H$_2$, and so to effectively suppress Pop III star formation.  

Figure 14 highlights the effects that the ionizing radiation have on the state of the gas (see also Figs. 8-10). 
The more massive BHs, initially emitting more ionizing photons, have a stronger dynamical impact on the gas, as shown 
in the left panels, causing more disruption of the dense gas surrounding the BH.  In turn, the lower gas densities 
lead to lower recombination rates in the central dense regions, allowing the ionizing radiation is to 
propagate further and heat the gas at larger distances from the BH.  As well, the limited amount 
of ionizing radiation generated by the 10$^4$ M$_{\odot}$ BH is evident in the bottom panels, as the H~{\sc ii} region
is bottled up at small radii and thus the radiation affects only a small region surrounding the black hole.

Also evident in Fig. 14 is the anisotropic expansion of the H~{\sc ii} regions surrounding the BHs, which arise from 
variations in the density of the gas in the center of the halo.  Even a relatively small difference in the density of the 
gas encountered by the ioinization front as it moves out from the central BH can translate into a large difference in the 
recombination rate, which scales as density squared, and so in the rate at which the ionizing photons are absorbed by the gas.  
Therefore, as the ionization front encounters a large range of densities of the gas at a give radius (see Figs. 8-10) as it expands outward, 
in directions in which the density is lower it extends out further than in directions in which the density is higher.  This 
is a generic feature of H~{\sc ii} regions formed in a cosmological density field, and has also been found for the case of H~{\sc ii} regions generated by 
stars (see e.g. Alvarez et al. 2006; Abel et al. 2007).

\subsection{Escape Fraction of Ionizing Photons}
Intertwined with the emission of ionizing radiation from the accretion disk and the hydrodynamic response of the gas to the radiation is 
the fraction of ionizing photons which are able to escape from the host halo.  To calculate this fraction, at every timestep we add up the total 
number of photons which propagate beyond 1 physical kpc from the BH; as this is well beyond the virial radius of the host halo ($\sim$ 600 pc) where 
the gas densities approach those of the intergalactic medium (IGM) (see Fig. 14), this provides a good estimate of the fraction of ionizing 
photons that can escape and contribute to the reionization the general IGM.   

Figure 15 shows the escape fraction $f_{\rm esc}$ of ionizing photons for the two most massive BHs; no ionizing radiation escapes the host halo in the case of the 
10$^4$ M$_{\odot}$ BH, the H~{\sc ii} region surrounding this BH being confined to the central regions of the halo, as shown in Figs. 10 and 14.  We find that only a small
fraction, never exceeding $\sim$ 0.1, of the ionizing photons are able to escape the host halo; indeed, the time-averaged escape fraction
is $\la$ 10$^{-2}$, for all the cases we studied. 

We emphasize, however, that the escape fractions that we calculate here do not account for photons with energies higher than those emitted from the thermal multi-color disk from 
which we derive our source spectrum.  In particular, X-ray photons with energies e.g. $\ga$ 1 keV, such as may be emitted in a non-thermal power law spectrum, could easily escape 
from the host halo and contribute to the reionization and photoheating of the IGM (e.g. Ricotti \& Ostriker 2004; Kuhlen \& Madau 2005; Volonteri \& Gnedin 2009).  
However, even accounting for the long mean free path of such higher energy emission, we expect that the majority of ionizing photons are still emitted from the thermal disk that we consider 
(see the observed spectra in e.g. Farrell et al. 2010), and hence that the error in our calculation of the escape fraction is not large. 

While the ionizing photon production rate (see Fig. 13) is, at least initially, comparable to what could be produced by a cluster of very massive Pop~III stars in the center of a 
similar halo, the escape fraction that we find is much lower than in the case of such a stellar cluster, which can approach unity
(Johnson et al. 2009; see also Wise \& Cen 2009).  This is a product of the self-regulation of the radiative output from accreting BHs, as the high energy radiation they emit
heats and disperses the gas, in turn limiting the accretion rate and the emission of radiation. In the case of stellar clusters, once they are formed 
their radiative output is not coupled to the properties of the surrounding gas, allowing them to emit copious amounts of ionizing radiation even after dispersing the gas 
surrounding them; this leads to generally higher escape fractions, one of the most extreme cases being that of 
massive Pop~III stars formed in minihaloes (see e.g. Whalen et al. 2004; Alvarez et al. 2006).

\begin{figure}
\vspace{2pt}
\epsfig{file=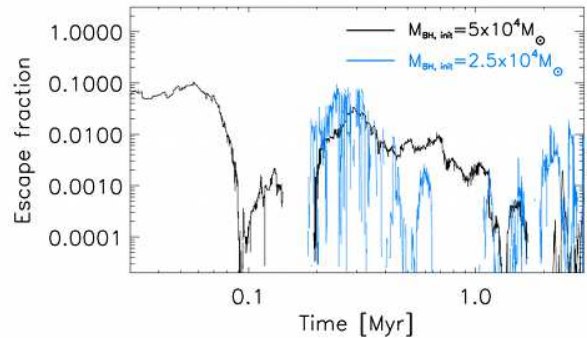,width=8.4cm}
\caption{The escape fraction $f_{\rm esc}$ of ionizing photons from the halo hosting the accreting BH.  
The escape fraction is always zero for the 10$^4$ M$_{\odot}$ BH case and so, for clarity, is not shown.  
The escape fraction is higher for the more massive BHs, in general, although for none of the cases does the escape fraction exceed $\sim$ 0.1.  
Indeed, the overall fraction of ionizing radiation escaping to reionize the the IGM is very low, the time-averaged escape fraction being well below that maximum value.     
}
\end{figure}

\begin{figure}
\vspace{2pt}
\epsfig{file=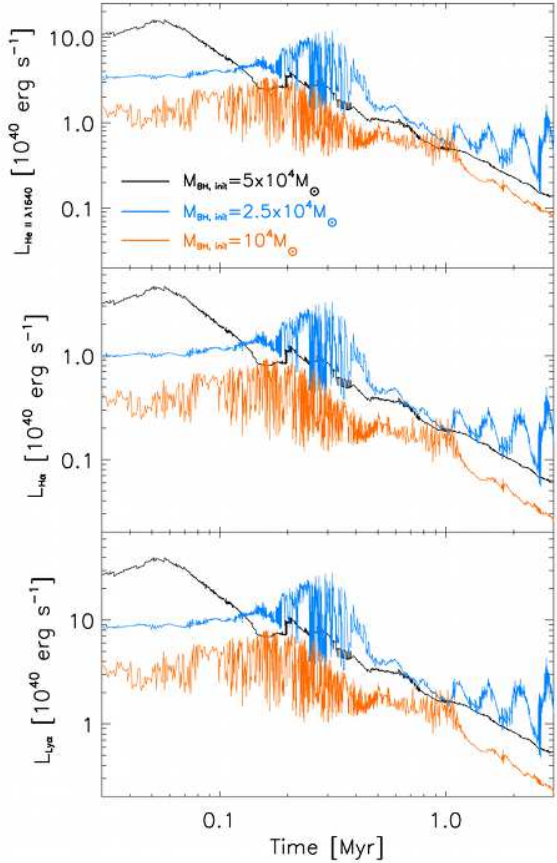,width=8.4cm}
\caption{Luminosity emitted from the ionized gas surrounding the accreting BHs in three prominent recombination lines which may be observed:  He~{\sc ii} $\lambda$1640 ({\it top panel}), H$\alpha$ ({\it middle panel}), and Ly$\alpha$ ({\it bottom panel}).  In part owing to the low escape fraction of ionizing photons, these luminosities are roughly proportional to the accretion rate (see Fig. 12).  Accordingly, note that while the luminosity is initially higher for a more massive BH, this is not always true, the luminosities in all cases generally decreasing in time, but at differing rates.  
}
\end{figure}

\section{An Observational signature of black hole accretion}
In this Section, we investigate a key signature of the observable radiation emitted from haloes hosting BHs accreting primordial gas in the early Universe: the 
strength of the emission in several prominent recombination lines. 

\subsection{Recombination Line Luminosities} 
The luminosity emitted in a given recombination line is given by (e.g. Schaerer 2002; Osterbrock \& Ferland 2006)

\begin{equation}
L = \alpha (1-f_{\rm esc}) Q \mbox{\ ,}
\end{equation}
where $\alpha$ is the energy emitted in the line for each recombination, $Q$ is the number of photons emitted per unit time that ionize the chemical species which produces the emission, 
and $f_{\rm esc}$ is the fraction of these photons that escape the halo.  For
our case, we use either $Q_{\rm HII}$ or $Q_{\rm HeIII}$, shown in Fig. 13.  Also, we do not differentiate between the escape fraction of H~{\sc i}- and He~{\sc ii}-ionizing photons, as we find that 
the H~{\sc ii} and the He~{\sc iii} regions surrounding the BH are almost completely overlapping, due to the high ratio of $Q_{\rm HeIII}$ to $Q_{\rm HII}$; in any case, the escape 
fraction is so low as to not be critical for our estimate of the recombination line luminosities.

The luminosities that we find, using the values for $\alpha$ given in Schaerer (2002), for Ly$\alpha$, H$\alpha$, and He~{\sc ii} $\lambda$1640 are shown in Fig. 16. 
Due to the low values of $f_{\rm esc}$, the luminosities we find here mirror almost exactly the ionizing photon production rates shown in Fig. 13; in turn, variations
in the luminosities also closely track those in the BH accretion rate.

\begin{figure}
\vspace{2pt}
\epsfig{file=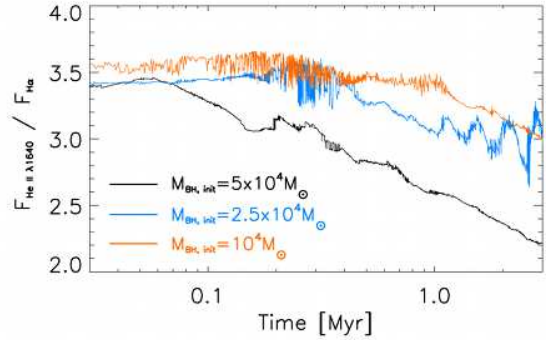,width=8.4cm}
\caption{The ratio of the total flux in He~{\sc ii} $\lambda$1640 to that in H$\alpha$, as a function of time since the formation of the BH, for each of the BH masses in our study.  Even for very massive Pop III stars, this ratio is $\le$ 2 (Johnson et al. 2009); therefore, one distinctive observational signature of accretion onto BHs formed by direct collapse is a high ratio of these fluxes, although this ratio varies in time for a given initial BH mass.  Interestingly, it is higher for the cases of the less massive BHs, owing to the harder spectrum of radiation emitted from their accretion disks, for a given accretion rate (see Fig. 7).  
}
\end{figure}

The luminosities in He~{\sc ii} $\lambda$1640 relative to those in H$\alpha$ or Ly$\alpha$, for all three cases, are extremely high.  The ratios of the observable flux in 
He~{\sc ii} $\lambda$1640 to that in H$\alpha$, to which the IGM is optically thin even before reionization is complete, are shown in Fig. 17. 
In all cases, this flux ratio exceeds $\sim$ 2, substantially higher than that expected even for very massive Pop~III stars (i.e. $\ga$ 100 M$_{\odot}$; see e.g. Johnson et al. 2009).  This suggests that a distinct observable signature of BHs accreting primordial 
gas in the early Universe could be such extremely high flux ratios, ultimately due to the very hard spectra emitted from the BH accretion disk.

\subsection{Prospects for Detection}
To evaluate the prospects for detecting this observational signature of BH accretion in the early Universe using the JWST, we estimate the total flux in the He~{\sc ii} $\lambda$1640
line as 

\begin{eqnarray}
F_{\rm 1640} &   =   &  \frac{L_{\rm 1640}}{4 \pi D_{\rm L}^2}  \nonumber \\ 
               & \sim & 10^{-20} \; {\rm erg} \; {\rm s}^{-1} \; {\rm cm}^{-2}  \left(\frac{L_{\rm 1640}}{10^{40} \; {\rm erg} \; {\rm s}^{-1}}\right) \left(\frac{1+z}{10}  \right)^{-2}  \mbox{ \ }
\end{eqnarray}
where $L_{\rm 1640}$ is the luminosity emitted in the line and $D_{\rm L}$($z$) is the luminosity distance to redshift $z$.  

The flux limit for 3 $\sigma$ detection of the He~{\sc ii} $\lambda$1640 line emitted from a source at $z$ $\sim$ 10, with 100 hours of observation by the Near-Infrared Spectrograph (NIRSpec) that 
will be aboard the JWST\footnote{http://www.stsci.edu/jwst/science/sensitivity/} is of the order of 10$^{-19}$ erg s$^{-1}$ cm$^{-2}$.
Therefore, at the peak luminosities in this line, shown in Fig. 16, emission due to both the 2.5 $\times$ 10$^4$ and the 5 $\times$ 10$^4$ M$_{\odot}$ BHs would be marginally detectable at $z$ $\la$ 10.
Given a small optical depth to Ly$\alpha$ through the IGM, or to sufficiently high velocity of the emitted gas relative to the IGM (see Dijkstra \& Wyithe 2010), 
the Ly$\alpha$ emission due to these BHs could also be detectable at the same redshift.  However, due to lower instrument sensitivity at the 
relevant wavelengths and to its lower luminosity, the flux in the H$\alpha$ line would be too low for detection.  

We note, however, that the accretion rates needed for these recombination lines to be bright enough for detection may be only short-lived, and the average 
accretion rates that we find are much lower than those required. That said, we have only tracked the accretion of gas for 3 Myr after the formation of the BHs; 
at later times, with the growth of the host halo by continuous accretion and mergers, the BHs could accrete gas at high rates again.  Also, as we have shown here, 
the luminosity of the radiation emitted due to the accretion process is sensitive to the mass of the BH, such that the emission due to more massive BHs than considered here will likely be more easily detected.

Another important question with regard to detecting BHs formed by direct collapse is the cosmological environment in which they are most likely to form.  
Given the high LW background flux needed for the formation of such BHs, 
Dijkstra et al. (2008) argue that the host haloes of these BHs may lie within $\sim$ 10 physical kpc of another halo with mass $\sim$ 10$^9$ M$_{\odot}$.  
Thus, it may the case that they are found very close to other, more luminous star-forming galaxies.
However, this distance and neighboring halo mass are sensitive to the level of the LW background that is required, 
and this could be significantly lower than the J$_{\rm LW}$ = 10$^3$ value used both in the present work and in the fiducial model of Dijkstra et al. 
(see Shang et al. 2010), in which case they may form further from $\sim$ 10$^9$ M$_{\odot}$ haloes or near less massive neighboring haloes.

\section{Implications for galaxy and black hole coevolution}
Numerous observations of high redshift galaxies suggest that the central BHs they host are more massive than would be expected from the observed BH - host galaxy 
relations at lower redshift (e.g. Walter et al. 2004; McLure et al. 2006; Peng et al. 2006; Bennert et al. 2010; Merloni et al. 2010; Greene et al. 2010; Decarli et al. 2010).  
It is interesting to note that the expected mass range of BHs (i.e. 10$^4$ - 10$^5$ M$_{\odot}$) formed by direct collapse in atomic-cooling haloes is at least qualitatively in line with this trend.
For instance, Ferrarese (2002) suggests a relation between the host halo mass and BH mass of $M_{\rm BH}$/$M_{\rm halo}$ $\la$ 10$^{-5}$ for $M_{\rm halo}$ $\la$ 10$^{12}$ M$_{\odot}$, 
whereas the ratio we have in the present work is two orders higher than this at the time of BH formation.  Furthermore, from their simulations of BH growth, Booth \& Schaye (2010) find that $M_{\rm BH}$ $\sim$ 10$^8$ M$_{\odot}$ ($M_{\rm halo}$ / 10$^{13}$ M$_{\odot}$)$^{1.55}$; hence, the $\ga$ 10$^4$ M$_{\odot}$ BHs
that we consider forming in $\la$ 10$^8$ M$_{\odot}$ haloes would lie roughly four orders of magnitude above the relation that these authors find.

While we track the growth of these black holes for only 3 Myr in the present work, we can extrapolate our findings to roughly estimate how the BH to host halo mass ratio may evolve at later times.
We find in our simulation that the host DM halo grows at a rate of $\sim$ 0.5 M$_{\odot}$ yr$^{-1}$ during the time that the BHs are accreting; using an average BH accretion rate from Fig. 12, $\sim$ 10$^{-5}$ M$_{\odot}$ yr$^{-1}$, along with this rate, we find that the ratio $M_{\rm BH}$/$M_{\rm halo}$ may drop to $\sim$ 10$^{-4}$ over $\sim$ 10$^9$ yr, still well above the relations suggested by both Ferrarese (2002) and Booth \& Schaye (2010).  However, with continued merging of the host halo with other haloes hosting BHs, such overly massive BHs could still eventually 
end up on the present-day observed scaling relations (Hirschmann et al. 2010; see also Volonteri \& Natarajan 2009).   

Due to the strong radiative feedback from the BHs, we find that star formation is not likely to take place in the central dense regions of the host halo, unless the BH accretion rate drops even further 
than we find in our simulations or the host halo is able to grow massive enough that some gas can become shielded from the photodissociating radiation emitted from the accretion disk 
(see also Oh \& Haiman 2002).  If star formation does occur, the accretion rate may be boosted by the transfer of angular momentum outward due to supernova explosions, thereby driving 
more gas to the center of the halo (see Chen et al. 2009; Kumar \& Johnson 2010).  If star formation does not occur, 
then BHs formed by direct collapse may remain unaccompanied by stars in their host halo, at least until the host halo merges with 
another halo in which stars may have already formed.  Indeed, it may be that the host halo ultimately merges with the halo or haloes in which reside the sources of radiation that are 
responsible for the high background LW radiation field that led to the formation of the BH to begin with (see e.g. Dijkstra et al. 2008).

\section{summary and discussion}
We have presented cosmological simulations of the formation and growth of black holes formed by direct collapse 
in an atomic-cooling halo at $z$ $\sim$ 15.  In these simulations, 
the accretion of gas and the detailed interaction of the gas with the high energy radiation emitted in the accretion process were modeled self-consistently, allowing us 
to arrive at a number of novel conclusions, which are summarized below.  

During the growth of the host halo prior to the formation of the BH by direct collapse, we find little evidence for supersonic 'cold flows' into the center of the halo, 
due to the inability of the gas to cool under the influence of the elevated photodissociating background radiation field, 
which is a likely prerequisite for this scenario of BH formation. 
Nonetheless, in the absence of radiative feedback, we find an upper limit to 
the rate at which the warm ($\ga$ 10$^3$ K) gas accretes onto a central supermassive star of the order of 0.1 M$_{\odot}$ yr$^{-1}$, 
high enough for it to grow to a mass of $\ga$ 10$^5$ M$_{\odot}$ before it collapses to form a BH.

With the formation of the BH and a hot accretion disk, photoheating of the gas in the host halo, as well as the radiative pressure to which it is subjected, 
lead to its expansion, which in turn leads to accretion rates which generally fall with time.  While in all three cases of initial BH mass that we simulate the accretion rate initially approaches the Eddington limit, within the 3 Myr of our simulations the accretion rate drops by at least an order of magnitude.  The time-averaged accretion rates 
that we find are $\la$ 10 percent of the Eddington rate.  

In part due to these low accretion rates, the time-averaged ionizing photon escape fractions are very low ($\la$ 0.01); this 
suggests that BHs formed by direct collapse contribute little to the reionization of the IGM in the vicinity of their host haloes.
These low escape fractions, in combination with the hard spectra emitted from the accretion disks, 
lead to an extremely high ratio of the nebular recombination radiation emitted in He~{\sc ii} 
$\lambda$1640 to that emitted in H$\alpha$ or Ly$\alpha$.  This is a distinctive observational signature of BH accretion in the early Universe, 
which may be detected by the JWST.  Another potential observational signature is that BHs formed by direct collapse may not be accompanied 
by stars formed in their host halo, due to the strong radiative feedback from the accretion process.
Finally, the large ratio of the masses of these BHs to the masses of their host haloes may provide some insight into the origin 
of observed BHs at high redshift which have higher masses, for a given host halo mass, than is observed in the local Universe.

We emphasize that the accretion rates that we report here are likely to be upper limits, for two reasons.  The first is that we do not consider the radiative feedback from 
the supermassive star that may form as the immediate predecessor to the massive black hole (see e.g. Bromm \& Loeb 2003; Volonteri \& Begelman 2010), which may 
act to decrease the accretion rate of the BH once it forms by driving gas away from the center of the halo (see e.g. Johnson \& Bromm 2007; Alvarez et al. 2009).
Secondly, as mentioned in Section 4.1, not all of the gas that is accreted at the scale of the Bondi radius is likely to be accreted onto the BH, as we have assumed.  

However, we also note that at scales below those that we resolve in our simulations the radiative feedback on the gas may be limited by 
shielding of the gas in the shadow of the accretion disk of the BH, as mentioned in Section 4.3.  Such a reduction in the radiative feedback on the 
accreting gas would likely allow for somewhat higher accretion rates.  Thus, while both the strength of the radiative feedback and the accretion rates that we find here
may be overestimates separately, how each of these would change in response to more detailed modelling of the other remains to be fully understood.
To better estimate the accretion rates of BHs in the early Universe, higher resolution cosmological simulations which follow the formation and evolution of 
the accretion disk around the BH would be of great interest.

The sub-Eddington accretion rates that we find may pose a significant challenge to models that call for the formation of $\ga$
10$^9$ M$_{\odot}$ BHs by $z$ $\sim$ 6 by accretion of gas onto black holes formed by direct collapse at 10 $\la$ $z$ $\la$ 15, 
and also for models which envision similarly massive BHs as forming from Pop III stars fueled by dark matter annihilation 
(see e.g. Freese et al. 2010).  It is important to note, though, that the single host halo that we have considered in this study represents only 
a 3-$\sigma$ fluctuation of the cosmological density field, while the most massive BHs at $z$ $\sim$ 6 likely reside in much more rare, 
faster growing haloes (e.g. Barkana \& Loeb 2001).  Therefore, while radiative feedback is certain to hinder the accretion of gas 
even in such rare overdense regions to some extent, the faster growth of the host halo would undoubtably lead to generally higher 
average accretion rates (see Li et al. 2007).  However, such a rare, fast growing halo would likely 
have a progenitor which hosted the formation of a Pop III star at very early times (i.e. at $z$ $\ga$ 30), before a substantial LW 
background radiation field could have been established (see e.g. Johnson et al. 2008).  While this would likely preclude 
the formation of a BH by direct collapse in the same halo at a later time, such Pop~III relic BHs 
could well be the seeds for the most massive BHs observed to power high redshift quasars.

\section*{Acknowledgements}
We are grateful to the support staff of the SFC supercomputer cluster at the Rechenzentrum Garching of the Max Planck Society, 
on which the simulations presented here were carried out.  JLJ would also like to thank Mitch Begelman, Dominik Schleicher, and Zolt{\' a}n Haiman for helpful discussions, as well as the anonymous referee for comments which improved the clarity of this work.

\end{document}